\documentclass[a4paper,fleqn,usenatbib]{mnras}

\usepackage{newtxtext,newtxmath}

\usepackage[T1]{fontenc}
\usepackage{ae,aecompl}

\usepackage{graphicx}	
\usepackage{amsmath}	
\usepackage{amssymb}	

\title[quenched massive spiral galaxies]{What has quenched the massive spiral galaxies?}

\author[Y. Luo et al.]{
Yu Luo,$^{1}$\thanks{E-mail: luoyu@pmo.ac.cn}
Zongnan Li,$^{2}$
Xi Kang,$^{3,1}$\thanks{E-mail: kangxi@zju.deu.cn}
Zhiyuan Li$^{2}$
and Peng Wang$^{4}$
\\
$^{1}$Purple Mountain Observatory, No. 10 Yuanhua Road, Nanjing 210033, China\\
$^{2}$School of Astronomy and Space Science, Nanjing University, China\\
$^{3}$Zhejiang University-Purple Mountain Observatory Joint Research Center for Astronomy, Zhejiang University, Hangzhou 310027, China\\
$^{4}$Leibniz-Institut fu\"r Astrophysik Potsdam (AIP), An der Sternwarte 16, D-14482 Potsdam, Germany
}

\date{Accepted XXX. Received YYY; in original form ZZZ}

\pubyear{2020}

\begin{document}
\label{firstpage}
\pagerange{\pageref{firstpage}--\pageref{lastpage}}
\maketitle

\begin{abstract}
Quenched massive spiral galaxies have attracted great attention recently, as more data is available to constrain their environment and cold gas content. However, the quenching mechanism is still uncertain, as it depends on the mass range and baryon budget of the galaxy. In this letter, we report the identification of a rare population of very massive, quenched spiral galaxies with stellar mass $\gtrsim10^{11}{\rm~M_\odot}$ and halo mass $\gtrsim10^{13}{\rm~M_\odot}$ from the Sloan Digital Sky Survey at redshift $z\sim0.1$. Our CO observations using the IRAM-30m telescope show that these galaxies contain only a small amount of molecular gas. Similar galaxies are also seen in the state-of-the-art semi-analytical models and hydro-dynamical simulations. It is found from these theoretical  models that these quenched spiral galaxies harbor massive black holes, suggesting that feedback from the central black holes has quenched these spiral galaxies. This quenching mechanism seems to challenge the popular scenario of the co-evolution between massive black holes and massive bulges.

\end{abstract}




\section{Introduction}

Galaxies are complex ecosystems composed of dark matter and multi-phase baryonic components, where gas accretion, cooling, star formation and feedback occur on different physical scales \citep{Tumlinson2017}.
Overall, galaxies display a clear bi-modal distribution in the color-stellar
mass or color-morphology diagram, in the sense that massive galaxies are
mostly red and bulge-dominated, while low mass ones have blue colors with disky
morphologies \citep[e.g.][]{Blanton2003}. However, there is a rare population of
galaxies that are quenched in their star formation albeit with disky
morphologies. Quenched spiral galaxies have been noticed for a long time \citep[e.g.][]{van den Bergh1976,Couch1998} and received more attention in recent years with the advent of large sky surveys \citep[e.g.][]{Masters2010, Rowlands2012, Fraser-McKelvie2018, Hao2019, Mahajan2020}. However, most existing studies have focused on observational facts rather than the quenching mechanism, in particular in the case of massive, group-central spiral galaxies.  

To quench star formation in a massive galaxy, all channels leading to the
accumulation of copious molecular gas have to be shut down \citep{Man2018}.
Unlike satellite galaxies whose cold gas can be relatively easily removed by
environmental effects, quenching of central spiral galaxies is more complicated.
Quenching does not necessarily mean that cold gas is absent. For example,
\citet{Zhang2019} showed that quenched spiral galaxies with stellar mass in the range of
$10^{10.6}~{\rm M_{\odot}} < M_{\ast} < 10^{11}~{\rm M_{\odot}}$ at
$0.02<z<0.05$ have HI mass around $10^{10}~\rm M_{\odot}$, which is
systematically higher than that of elliptical galaxies with similar stellar masses. This
result suggests that the prevalence of diffuse HI gas is the main reason for
preventing continuous star formation. However, it is not without controversy, for
instance, \citet{Cortese2020} recently reported that the bulk of passive disk
galaxies in the GALEX Arecibo SDSS Survey \citep[xGASS,][]{Catinella2018} are still HI-poor across the stellar mass range of $ 10^9<M_*/{\rm M_\odot}<10^{11}$. 

In addition to the uncertain existence of cold gas (either molecular or hydrogen) in quenched central spiral galaxies, the need for quenching depends on the halo mass and the baryon budget in the galaxy. For example, \citet{Li2019} studied a massive isolated spiral galaxy, NGC5908, which is considered inactive in star formation. By measuring various baryonic components, they concluded this galaxy is ``missing baryon'', not completely quiescent, and probably at an early evolutionary stage after a fast growth stage. This kind of baryon-deficient galaxies may be the progenitors of the local massive quenched spiral galaxies. This scenario circumvents extra feedback to suppress gas cooling, but it has to face the missing baryon problem. However, \citet{Posti2019} found that most massive central spirals with stellar masses below $\rm 10^{11} ~M_\odot$ are living in medium-mass haloes (a few times $10^{12}~\rm M_{\odot}$) where almost all the halo gas has cooled and turned into stars in the disk. These results indicate that the observational uncertainty in the halo mass may affect the explanation.

Overall, from the observational aspects, the formation route of massive quenched
spiral galaxies is still unclear. Nevertheless, for galaxies in massive haloes
(more than $\rm 10^{13}M_{\odot}$), most baryon is in the hot gaseous halo
with a quasi-universal fraction \citep[e.g.][]{Wang2017, Davies2020} .
In this mass regime, unless cooling of the hot gas is suppressed, the central
galaxy will get replenishment of its cold gas for continuous star
formation. A few mechanisms have been proposed to suppress gas cooling, among
which AGN heating is the most common solution
\citep{DiMatteo2005, Croton2006, Cattaneo2009}. Powerful AGN feedback requires
the existence of a massive black hole. However, according to the bulge-black
hole co-evolution scenario, a spiral galaxy without a prominent bulge is unlikely
to host a massive black hole. This poses a challenging question: what has quenched these massive central spiral galaxies? Based on the $\Lambda$CDM cosmology, the state-of-the-art semi-analytical models and hydrodynamical simulations (e.g., L-Galaxies  \citep{Guo2011, Henriques2015}; EAGLE simulation \citep{Schaye2015, Crain2015}; Illustris-TNG simulation \citep{Pillepich2018, Nelson2018, Springel2018}) have successfully reproduced many observations. It is then interesting to see whether they can predict the existence of quenched massive spiral galaxies.

In this letter, we report our investigation of a rare population of massive quenched central spiral galaxies, which are identified from the Sloan Digital Sky Survey (SDSS). Among them, we have selected four galaxies to probe the amount of molecular gas with the IRAM-30m telescope. By investigating the existence of such galaxies in both the state-of-the-art semi-analytical model and hydro-dynamical simulations, we propose the most plausible formation scenario.  

\section{Sample selection}
\begin{figure}
\centering
\includegraphics[scale=0.7]{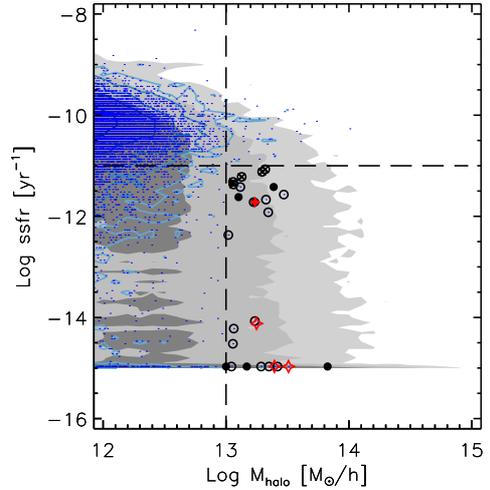}
\caption{The ssfr-halo mass distribution of central spiral galaxies in the SDSS. The dark and grey shadow shows the ssfr-halo mass distribution of all the central galaxies in our SDSS catalog. The blue contours show those 16335 central galaxies with fracDev$_{r}<0.1$. The rest symbols (circles and stars) are our 27 quenched massive central spiral sample. Among them, 6 galaxies showing AGN activity according to the BPT diagram are marked by the filled symbols; 5 galaxies that are above the quenched threshold of \citet{Trussler2020} are marked with open circles plus crosses; the 4 galaxies observed by IRAM-30m are marked by red stars.The horizontal and vertical dashed lines are our criteria to select quenched and massive galaxies.}
\label{Fig:ssfr-mass}
\end{figure}

We identify a rare population of quenched massive spiral galaxies from the publicly available galaxy group catalogue \citep{Yang2012}
constructed from the SDSS Data Release 7 \citep{Abazajian2009}.  
Firstly, we select disk-dominated central galaxies with the criterion of fracDev$_r < 0.1$, where fracDev$_r$ is the the coefficient of the de Vaucouleurs component, which is a good indicator of the bulge-to-total mass ratio. These galaxies are then separated into star-forming and quenched ones via the specific star formation rate ${\rm sSFR=SFR}/M_*$ with a threshold of $\rm 10^{-11}\rm yr^{-1}$. In this work, the sSFRs of galaxies are from \cite{Chang2015}, which combines optical (SDSS) and infrared (Wide-field Infrared Survey Explorer; WISE, \citealp{Wright2010}) photometry. 
At this point we have 16335 central spiral galaxies, most of which are star-forming galaxies (Fig.~\ref{Fig:ssfr-mass}). 
We then select those in massive haloes $M_{\rm halo} \geq 10^{13}h^{-1}\rm~M_{\odot}$, resulting in 72 disky central galaxies. Among them there are 27 quenched galaxies with stellar mass
around $2\times 10^{11}\rm~M_{\odot}$, which is larger than the quenched galaxies studied in some previous work
\citep[e.g.][]{Zhang2019}. For comparison, we also use a mass-dependent sSFR threshold  \citep{Trussler2020} to 
separate star-forming and quenched galaxies. 
It is found that the majority of these 27 galaxies are still classified as quenched;  
only 5 galaxies (marked with crosses in Fig.~\ref{Fig:ssfr-mass}) become star-forming,  
but they still belong to the green valley according to the \citet{Trussler2020} threshold.

These 27 quenched disky galaxies are all found in isolated environment: their distances to the nearest central galaxy are larger than 1 Mpc and their group richness is less than 5. It suggests that they are not living at the outskirt of nearby over-dense region where environmental quenching may be effective. In other words, these galaxies are quenched either because they lack new cold gas supply, or there is a reservoir of cold gas, but the star formation is suppressed by certain mechanism. Thus probing other forms of baryonic mass other than stellar mass is key to understand why these galaxies are quenched.

We have searched for signs of AGN activities from publicly available databases, in the radio (FIRST \citep{Becker1995}, NVSS \citep{Condon1998} and LOFAR \citep{vanHaarlem2013}) and X-ray (Chandra, XMM-Newton) bands, but found no counterpart for any of the sample galaxies. This is consistent with our finding from the BPT diagram that only 6 of the 27 galaxies show weak AGN activities (marked by the filled symbols in Fig.~\ref{Fig:ssfr-mass}). 

We have also looked for detections of HI or molecular gas using the ALFALFA \citep{Haynes2018}
and xCOLD GASS \citep{Saintonge2017} surveys. All but one galaxy in our sample have redshifts larger than 0.1, with one at $z\sim0.08$. Unfortunately, galaxies in the two surveys have redshifts lower than 0.06, far less than the redshift range of our samples. However, a recent study \citep{Zhang2019} shows that quenched spiral galaxies with stellar mass $10^{10.6}~{\rm M_{\odot}} < M_{\ast} < 10^{11}~{\rm M_{\odot}}$ at $0.02<z<0.05$ have HI mass around $10^{10}~\rm M_{\odot}$, systematically higher than that of elliptical galaxies with similar mass. The stellar mass-halo mass relation from the abundance matching \citep{Moster2013} suggests that the quenched galaxies of \citet{Zhang2019} are living in halos with $M_{\rm halo} < 3\times 10^{12}~\rm M_{\odot}$, well below our selection of halo mass above $10^{13}h^{-1}~\rm M_{\odot}$.

\section{CO observation}
\begin{figure}
\centering
\includegraphics[scale=0.25]{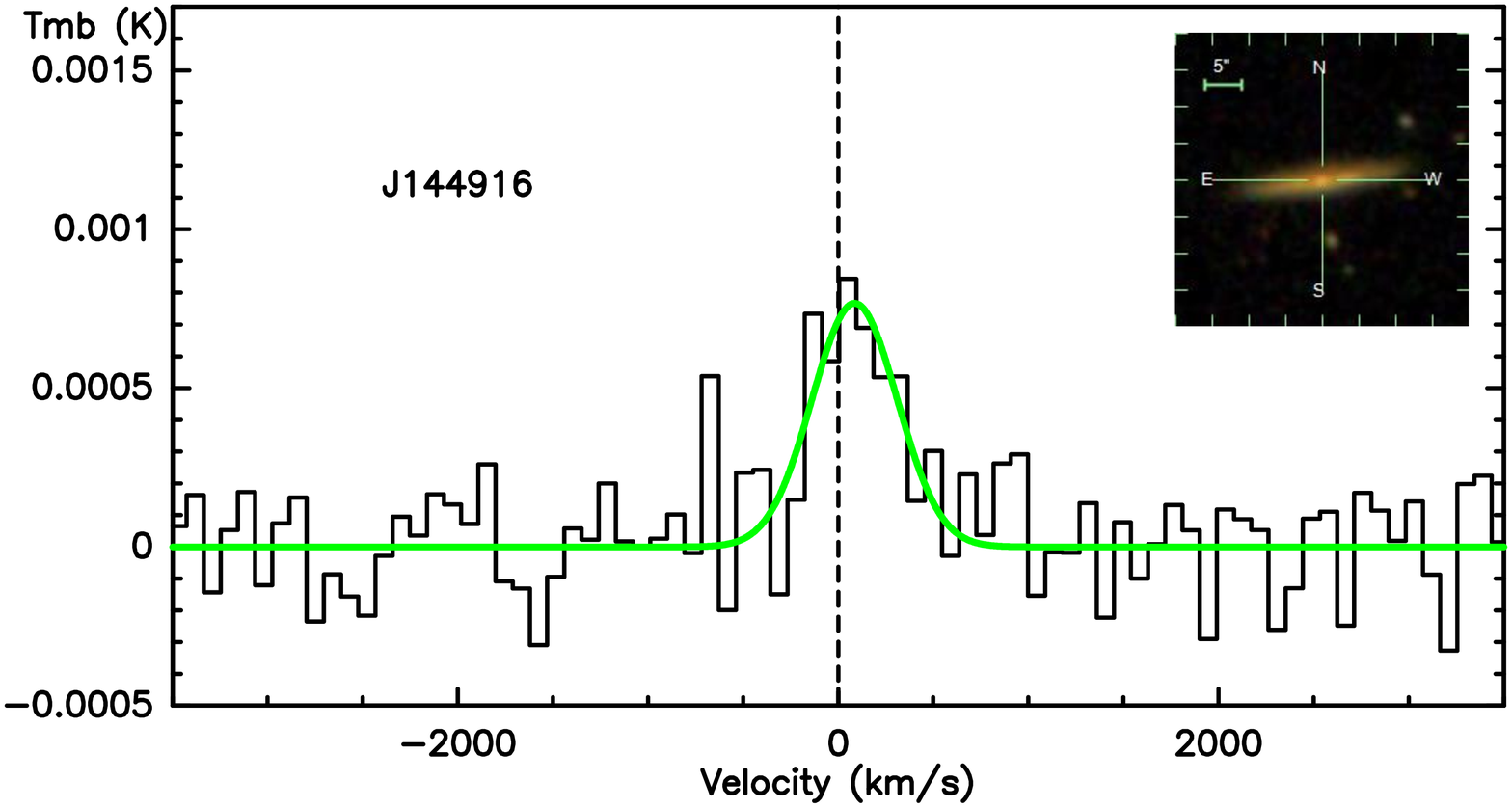}
\includegraphics[scale=0.25]{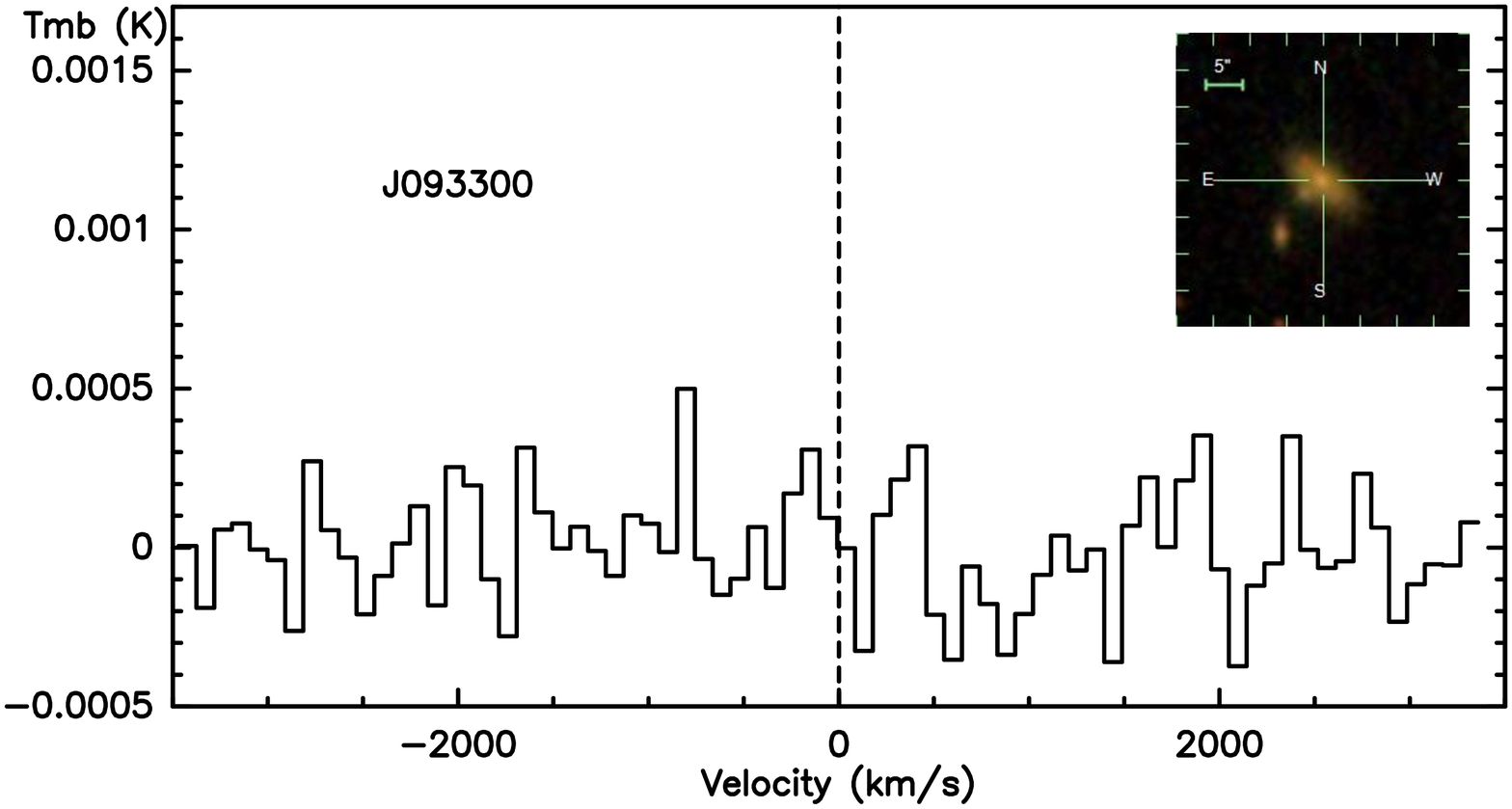}
\includegraphics[scale=0.25]{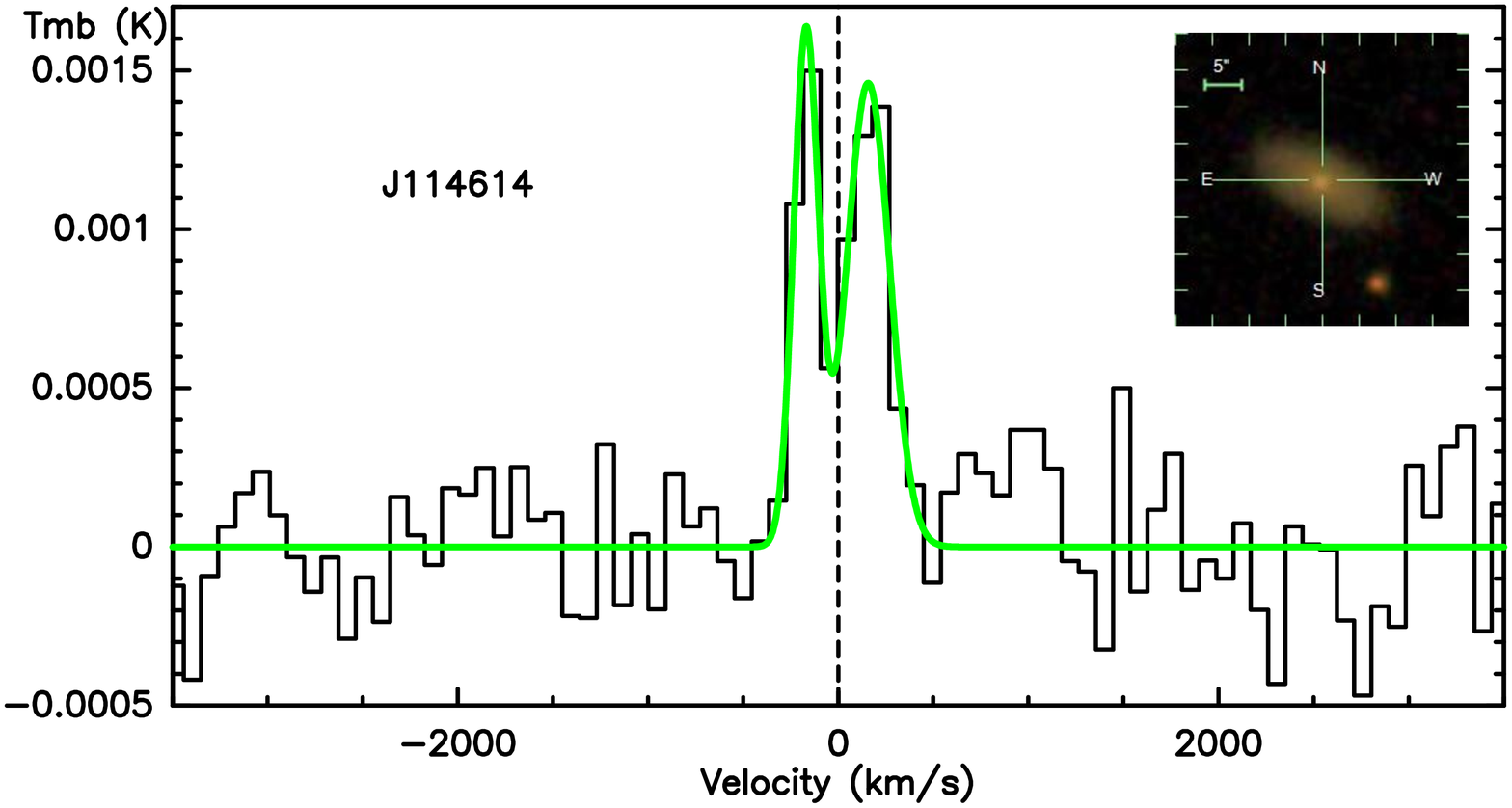}
\includegraphics[scale=0.25]{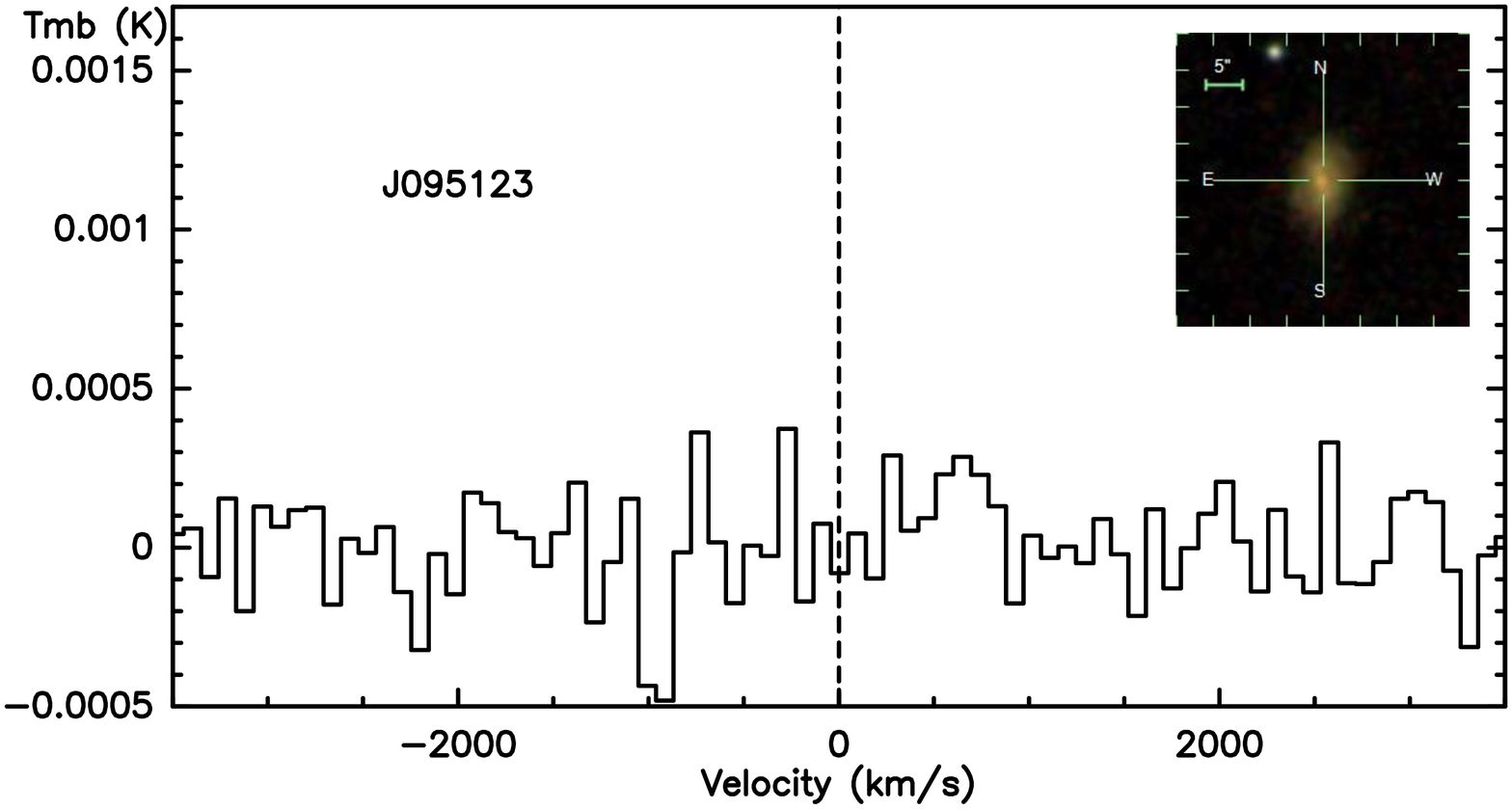}
\caption{The CO(1-0) spectra of four selected massive central spirals, the SDSS image of which is displayed in the insert. For the two galaxies with significant CO(1-0) emission, one or two Gaussians are applied to fit the spectra and the results are shown by the green curve.
The vertical dashed line indicates the corresponding redshift of each galaxy.}
\label{Fig:co}
\end{figure}

To put a rough limit on the amount of molecular gas in these quenched galaxies, we
select 4 galaxies from our sample of 27 galaxies to carry out CO observations using the IRAM-30m telescope. 
The four targets are chosen for an optimal combination of distance, sSFR and morphology.
The SDSS images of the four galaxies are shown in the insert of each panel in Fig.~\ref{Fig:co}, and their basic properties are given in Table~\ref{table:colog}.
The first galaxy, SDSS\,J144916, is an edge-on galaxy showing a clear disky morphology. The other three galaxies have extended disks and visible small bulges.

Detailed information of the CO observations and the fitting results of the derived spectra are given in Table~\ref{table:colog}. The individual CO spectra and the fitted Gaussian models (for detection only) are shown in Fig.~\ref{Fig:co}. Only two galaxies, SDSS\,J144916 and SDSS\,J114614, show a significant CO(1-0) line at a confidence level greater than 3$\sigma$.
The line intensity is converted from main beam temperature $T\rm_{mb}$ (Kelvin) to flux density S (Jansky) using $S/T\rm_{mb} \sim$ 5 Jy K$^{-1}$ at 100 GHz\footnote{http://www.iram.es/IRAMES/mainWiki/Iram30mEfficiencies}. The CO line luminosity $L\rm_{CO}$ is then calculated following the standard equation from \citet{Solomon1997}:
\begin{equation}
L\rm_{CO} = 3.25 \times 10^7 \it S\rm_{CO}\Delta \it V \nu\rm_{obs}^{-2} \it D\rm_L^2 (1+\it z)\rm^{-3}
\end{equation}
where $L\rm_{CO}$ is in K km s$^{-1}$ pc$^2$, the CO flux $S\rm_{CO}\Delta \it V$ obtained with Gaussian fitting, is in units of Jy km s$^{-1}$. The observing frequency $\nu_{\rm obs}$ is in GHz, and the luminosity distance $D\rm_L$ is in Mpc. We adopt the fiducial CO-to-H$_2$ conversion factor $\alpha\rm{_{CO(1-0)}} = 4.35~\rm M_{\odot}\, \rm{pc^{-2}\, (K\,km\,s^{-1})^{-1}}$, equivalent to $X\rm_{CO} = 2 \times 10^{20}~cm^{-2}$ (K km s$^{-1})^{-1}$ \citep[e.g.][]{Bolatto2013}  to convert CO luminosity to molecular gas mass. However, we note that this conversion factor is of large uncertainty, especially for the rare sample of massive quenched central spirals. For the two sources without a significant CO detection, we estimate the 3$\sigma$ upper limit by assuming a line width (full width at half maximum) of 500 km s$^{-1}$, similar to the observed values. 

All the four galaxies have amount of molecular gas less than $1.2\times 10^{10}~\rm M_{\odot}$.  The molecular gas fraction ($M_{\rm H_2}/M_*$) is $\lesssim$4\% in three galaxies and $\sim$8\% in the remaining galaxy (SDSS\,J114614).
For the two galaxies with CO detection, the circular velocity can be approximated by the 20\% peak line width of the spectra and certain inclination angle, which are 274.7 km~s$^{-1}$ and 283.2 km~s$^{-1}$ for SDSS\,J144916 and SDSS\,J114614, respectively. Assuming a typical NFW profile \citep{Navarro1997}, and the halo mass-concentration relation \citep{Prada2012}, we can estimated halo mass by fitting the circular velocity at the radius within which the CO is detected. We obtain similar halo mass of $10^{13.1}~\rm M_{\odot}$ for both two galaxies. Note that here, this mass is the lower limit as we assumed the size of  molecular gas extends to the stellar disk.
Adopting an equal upper limit of $10^{10}~\rm M_{\odot}$ for both HI and
molecular gas for our galaxies, the gas fraction is at most 10\%. Thus, we do
discover a rare population of quenched spiral galaxies with stellar mass around $2\times 10^{11}~\rm M_{\odot}$ and a small fraction of cold gas, living in haloes more massive than $10^{13}h^{-1}~\rm M_{\odot}$.

\section{Theoretical models for galaxy formation}
Both the semi-analytic models and hydro-dynamic simulations are powerful tools  to study galaxy formation and evolution in a cosmological context. We examine the state-of-the-art semi-analytic models and hydro-dynamic simulations to see whether they are able to predict the existence of galaxies analogous to our sample galaxies, and if any, to shed light on the cause of quenching in these massive spiral galaxies.

\subsection{Semi-Analytic Model}
Semi-analytical models (SAMs) are usually based on merger trees from N-body simulations and incorporate  phenomenological 
descriptions of various physical processes related to galaxy formation, such as cosmic 
reionization, hot gas cooling and cold gas infall, star formation and metal 
production, supernova feedback, gas stripping and tidal disruption of satellites, 
galaxy mergers, bulge formation, black hole growth and AGN feedback, etc. They 
can reproduce lots of observational facts such as the mass/luminosity function, 
the galaxy color-mass diagram and so on. L-Galaxies is one of the most successful semi-analytic galaxy formation models, which has been continuously developed by the 
Munich group in the last two decades. \citet[][here after H15]{Henriques2015} presented the latest L-Galaxies model, in which they have used the Markov Chain Monte Carlo (MCMC) method 
to search the parameter space, successfully fitting the evolution of stellar 
mass function and the overall fraction of red galaxies from $z=3$ to 0. 

Here we use the public catalog\footnote{http://gavo.mpa-garching.mpg.de/Millennium} of H15 implemented on the Millennium simulation. Following the sample selection from the SDSS, we 
select central galaxies at $z=0.1$ with $M_{\rm halo} >10^{13}h^{-1}~\rm M_\odot$ and 
${\rm B/T}=M_{\rm bulge}/M_*<0.1$ as our criteria for massive spirals. We also use
$\log \rm ssfr= -11~yr^{-1}$ to divide these massive central spirals into 
quenched and star-forming populations. Finally, we obtain 7024 quenched massive 
central spirals and 372 star-forming ones. There are more quenched 
galaxies in H15 than in the SDSS catalog, presumably due to the strong AGN feedback in the H15 model. 
After checking the different baryonic components of these galaxies in the SAM,
we find that the quenched sample has a higher hot gas fraction and a lower cold gas 
fraction than the star-forming ones. This clearly shows that quenching is directly due to the lack of gas cooling in these massive galaxies. Why, then, the hot gas did not cool? If there were no AGN feedback, both the quenched and star-forming samples would have a similar cooling rate (roughly higher than 
$\rm 200~M_{\odot} yr^{-1}$). The SAM suggests that AGN heating is the primary mechanism responsible for quenching the gas cooling and subsequent star formation.

\subsection{Hydro-dynamic simulation}
The Illustris-TNG project is a successful suite of large, cosmological 
magnetohydrodynamical simulations of galaxy formation which can 
reproduce many observational results, such as the stellar content distribution \citep{Pillepich2018}, color distribution \citep{Nelson2018}, matter and galaxy clustering \citep{Springel2018}.
The series of simulation include three box sizes: 50, 100, and 300 Mpc$^3$ with 3 different resolution respectively. In this work we use the largest box (TNG300-1) which is suitable for the study of massive galaxies.   

\begin{figure}
\centering
\includegraphics[scale=0.7]{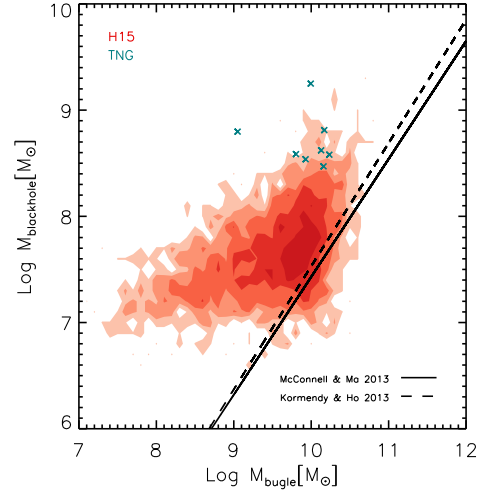}
\caption{Black hole mass-bulge mass relation for massive spiral galaxies from the H15
SAM and TNG simulation. 
The red contours and dark green crosses represent the H15 sample (with $ M_{\rm halo}>10^{13}h^{-1}{\rm M_\odot}$ and $ M_*>10^{11}\rm M_\odot$) and TNG sample (only with $M_*>10^{11}\rm M_\odot$) respectively. 
The solid line and dashed line are from the classical bulge - black hole mass relation of McConnell \& Ma (2013) and Kormendy \& Ho (2013).} 
\label{Fig:bh-mass}
\end{figure}

We select central galaxies with $M_*>10^{11}~ \rm M_\odot$ from TNG300-1 and decompose their stellar mass distribution into disk and bulge components. Finally we obtain only 8 massive central galaxies with B/T $<0.2$, all of which are quenched galaxies. 
To understand why these galaxies are quenched in the model,
we show in Fig.~\ref{Fig:bh-mass} the relation between black hole mass and bulge stellar mass from the SAM and Illustris-TNG, respectively. It can be seen that the predicted black hole mass of the eight quenched galaxies is above the empirical black hole mass-bulge mass relation \citep[e.g.][]{McConnell2013, Kormendy2013}, indicating that feedback from the central massive black holes have quenched their star formation.

\begin{table*}
\begin{center}
	\begin{tabular}{lcccc} 
		\hline
		\hline
                 Name & J144916.71+150700.1 & J093300.93+594420.4 & J114614.24+111501.7 & J095123.60+265643.9 \\
		\hline
RA(J2000) &  222.320 & 143.254 & 176.559 & 147.848 \\
DEC(J2000) & 15.117 & 59.739 & 11.251 & 26.946 \\
$z$ & 0.111 & 0.151 & 0.114 & 0.131 \\
$\log M_{\rm halo}({\rm{M_\odot}})^{(a)}$ & 13.40 & 13.55 & 13.66 & 13.29 \\
$\log M_*({\rm {M_\odot}})^{(b)}$ & 11.20 & 11.29 & 11.12 & 11.02 \\
$\log {\rm SSFR}({\rm yr^{-1}})^{(c)}$ & -14.12 & -14.97 & -14.97 & -11.72 \\
Frequency (GHz) & 103.75 & 100.15 & 103.48 & 101.92 \\
$T\rm{_{int}}^{(d)}$ (h) & 4.5 & 5 & 3 & 4.4 \\
${\rm RMS}^{(e)}$ (mK) & 0.15 & 0.17 & 0.22 & 0.15 \\
Velocity$^{(f)}$ (km s$^{-1}$) & 82.47 $\pm$ 34.21 &-&-169.40 $\pm$ 12.31&-\\
& & & 157.77 $\pm$ 17.30 \\
FWHM$^{(g)}$ (km s$^{-1}$) & 547.74 $\pm$ 100.90 &-& 155.36 $\pm$ 36.09&-\\
& & & 267.31 $\pm$ 43.26 \\
$T_{\rm mb,peak}^{(h)}$ (mK) & 0.93 $\pm$ 0.15 &-&1.95 $\pm$ 0.22&-\\
& & & 1.76 $\pm$ 0.22 \\
$I_{\rm CO}^{(i)}$ (K km s$^{-1}$) & 0.54 $\pm$ 0.07& $<$0.31 $^{(m)}$& 0.33 $\pm$ 0.06 &$<$0.28$^{(m)}$\\
& & & 0.51 $\pm$ 0.07 \\
log$L_{\rm CO}^{(j)}$ ($\rm L_\odot$) & 9.20 $\pm$ 0.06 & $<$ 9.24 & 9.41 $\pm$ 0.05 & $<$ 9.06\\
log$M_{\rm H_2,o}^{(k)}$ ($\rm M_\odot$) & 9.84 $\pm$ 0.06 & $<$ 9.88 & 10.05 $\pm$ 0.05 & $<$ 9.70 \\
log$M_{\rm gas,p}^{(l)}$ ($\rm M_\odot$) & 9.98 & 10.02 & 9.95 & 9.91\\
               \hline                
        \end{tabular}
\caption{Basic information and Observation log of our target galaxies. 
Note: $(a)$ Halo mass from Yang et al. group catalog. $(b)$ Stellar mass from Chang et al. catalog. $(c)$ SSFRs from Chang et al. catalog.  $(d)$ Integration time. $(e)$ The baseline RMS in 90 km s$^{-1}$ channel. $(f)$ Central velocity. $(g)$ Full-width at half maxima. $(h)$ Peak main beam temperature. $(i)$ Integrated intensity. $(j)$ CO luminosity. $(k)$ Observed molecular gas mass. $(l)$ Predicted cold gas mass from H15. $(m)$ Three Sigma upper limit by assuming a line width of 500 km s$^{-1}$.}\label{table:colog}
\end{center}
\end{table*}

\section{Conclusion and Discussion}
From the theoretical point of view, galaxies may still be quenched if they have plenty of cold gas but do not form star efficiently. I) dynamical processes related to a bar or bulge can stabilize the gaseous disk from fragmentation. This is ruled out for our observed sample since it requires the galaxies to possess significant bulges. II) There is a large diffuse HI gas which can not form stars directly in these quenched galaxies, as claimed by some recent work \citep{Zhang2019}. Nevertheless, this is refuted by \citet{Cortese2020} that the bulk of the passive disk galaxies in xGASS is still HI-poor across the stellar mass range $10^9<M_*/{\rm M_\odot}<10^{11}$. III) No quenching mechanism is needed if these massive central spirals are living in low mass haloes (a few times $10^{12}~\rm M_{\odot}$) where almost all gas has cooled and turned into stars in the disk, as recently found by \citet{Posti2019}. Both the latter two cases are possible as galaxies with masses lower than $10^{11}~\rm M_{\odot}$ are living in halos with masses lower than $3\times 10^{12}~\rm M_{\odot}$, where the rapid cold accretion along cosmic filaments is expected \citep{Dekel2009}. This cold gas can be kept in the form of HI gas around the central galaxy or transferred into stars once molecular gas formation becomes efficient. All these indicate that the halo mass is the key to infer the formation of these galaxies. For the massive ones ($> 10^{13}~\rm M_{\odot}$), extra strong feedback is necessary. 

The galaxies in our sample are more massive ($>10^{11}~\rm M_{\odot}$) and living in halos massive than $10^{13}h^{-1}~\rm M_{\odot}$, where the popular mass or halo quenching term applies. Whatever the mechanism behind the mass quenching, an energy source is needed to balance the cooling of the hot gas \citep{Man2018}. Our results from both the SAM and Illustris-TNG simulation suggest that the cooling from the hot gaseous halo in these quenched spiral galaxies are suppressed by massive black holes. Other than AGN heating, shock heating from infalling satellites may heat up the gas \citep{Khochfar2008}. In our sample, these galaxies are mostly isolated disk galaxies, thus mergers should not be important. After excluding these possibilities, we conclude that AGN heating, either still on-going on or being halted shortly before the current epoch, is the most plausible mechanism to suppress gas cooling in these galaxies. This conclusion calls for further careful search of massive black holes in our sample galaxies. The existence of such massive black holes will challenge current scenario of massive black hole formation in bulgeless galaxies, whereas their non-existence will put more tight constraints on the halo mass and other baryonic components.

\section*{Acknowledgements}
We thank the anonymous referee for useful comments. The Millennium Simulation data bases used in this paper and the web application providing online access to them were constructed as part of the activities of the German Astrophysical Virtual Observatory (GAVO). Based on observations carried out under project number 194-18 with the IRAM 30m telescope. IRAM is supported by INSU/CNRS (France), MPG (Germany) and IGN (Spain). Y.~L. and X.~K. thank partial support for this work by the 973 program (No.2015CB857003) and the National Natural Science Foundation of China (No.11825303, 118611311006, 11333008, 11703091). Z.-N.~L. and Z.-Y.~L. acknowledge support from the National Key Research and Development Program of China (2017YFA0402703).








%
%
%

\bsp	
\label{lastpage}
\end{document}